\documentclass[twocolumn,nofootinbib,amsmath,amssymb,aps,prd,balancelastpage,superscriptaddress]{revtex4-1}

\usepackage{color}
\usepackage[dvipsnames]{xcolor}
\usepackage[active]{srcltx}
\usepackage{amsmath,amsfonts,amssymb,amsthm,amstext,amscd,eucal,srcltx}
\usepackage{epsfig,graphicx,bm}
\usepackage{epstopdf, epsf}
\usepackage{dcolumn}
\usepackage{hyperref}
\usepackage{tensor}
\usepackage{lipsum}
\usepackage{appendix}
\usepackage{tikz}
\usepackage{caption}
\usepackage{subcaption}

\newcommand{\be}{\begin{equation}}
\newcommand{\ee}{\end{equation}}

\newcommand{\bse}{\begin{subequations}}
\newcommand{\ese}{\end{subequations}}
\newcommand{\bea}{\begin{eqnarray}}
\newcommand{\eea}{\end{eqnarray}}
\newcommand{\ba}{\begin{array}}
\newcommand{\ea}{\end{array}}
\newcommand{\bc}{\begin{center}}
\newcommand{\ec}{\end{center}}

\begin{document}
%\preprint{IPM/P-2012/009}
\vspace*{3mm}

\title{Black hole shadow and chaos bound violation in $f(T)$ teleparallel gravity}

\author{Andrea Addazi}
\email{addazi@scu.edu.cn}
\affiliation{Center for Theoretical Physics, College of Physics Science and Technology, Sichuan University, 610065 Chengdu, China}
\affiliation{Laboratori Nazionali di Frascati INFN, Frascati (Rome), Italy, EU}

\author{Salvatore Capozziello}
\email{capozziello@na.infn.it}
\affiliation{Dipartimento di Fisica "E. Pancini",  Universit\`{a} di Napoli "Federico II",
Compl. Univ. Monte S. Angelo Ed. G, Via Cinthia, I-80126 Napoli (Italy)}
\affiliation{and Istituto Nazionale di Fisica Nucleare  Sez. di Napoli, Compl. Univ. Monte S. Angelo Ed. G, Via Cinthia, I-80126 Napoli, Italy}
\affiliation{Scuola Superiore Meridionale, Largo S. Marcellino 10, I-80138 Napoli, Italy.}

%\date{\today}

\begin{abstract}

\noindent

We show that the so-called chaos bound, proposed by Maldacena, Shenker \& Stanford,  can be violated in $f(T)$ teleparallel gravity. 
In particular, it is possible to select  a new {\it gravitational Lyapunov parameter}, controlling {\it chaotization} of circular trajectories, in black hole photo-sphere, that can exceed the Maldacena, Shenker \& Stanford thermal bound.  This feature alters the standard quasi-normal modes and ringdown phase after black hole merging  with intriguing implications for future gravitational wave detections and black hole shadow measurements. It is a general characteristic for several gravitational theories beyond standard General  Relativity.

\end{abstract}

\maketitle

\section{Introduction}. 
\vspace{0.2cm}

One of the most intriguing features of several  black hole (BH) solutions in extended theories of gravity \cite{ETG}
is that they have a dynamical horizon even without in-falling of external matter or emission of Bekeinstein-Hawking radiation. 
This phenomenon seems to be almost ubiquitous in theories beyond Einstein's General Relativity (GR), including dilaton-gravity, $f(R)$ gravity and $f(T)$ teleparallel gravity \cite{A1,A2,A3,A4,A5,A6,A7,A8,A9,A10,A11,A12,A13,A14,A15,A16,Oiko,Manos,Addazi:2017vti,Addazi:2018pcc}. The impact of such a  fact on our conception of BH thermodynamics and entropy 
may be dramatically important. Indeed, in some cases, the Bekeinstein-Hawking radiation may 
be altered as a result of out of thermal equilibrium dynamics of BH horizon
\cite{Addazi:2016prb,Addazi:2017lat}.

Here, we want to show that also the so-called  {\it chaos bound}, proposed by {\it Maldacena}, {\it Shenker} and {\it Stanford} (MSS), is violated in extended theories like $f(T)$ teleparallel  gravity. 
In particular, a new Lyapunov exponent,  dubbed as {\it gravitational Lyapunov parameter} $\lambda_{G}$, emerges out from $f(T)$ gravity, competing with the standard Lyapunov parameter $\lambda_{0}$ of GR. According to MSS conjecture, the standard Lyapunov exponent is bounded
as $\lambda_{0}\leq 2\pi \beta^{-1}$ in every system with a temperature $\beta^{-1}$, posing 
a limit to the time growth rate of out of thermal equilibrium operators (OTOs) \cite{Maldacena:2015waa}. 
However, we will show that,  in $f(T)$ gravity,  $\lambda_{G}$ can be larger than $\lambda_{0}$ 
 and chaos in the system can propagate with higher efficiency than in GR, violating the MSS bound. 

As the standard Lyapunov parameter, 
also the gravitational Lyapunov one enters into BH 
Quasi-Normal-Modes (QNMs) as well as in 
butterfly effect of geodetic trajectories 
in the photo-sphere. 
We will show how chaotic geodetics
and QNMs are modified in $f(T)$ gravity
with important implications 
for BH shadow,  BH merging
and GW signals from the ringdown phase\footnote{See 
Refs.\cite{Gan:2021xdl,Khodadi:2021gbc,Okyay:2021nnh,Lin:2022ksb,Odintsov:2022umu,Chen:2022nbb,Jusufi:2022loj,Chen:2022qrw,Khodadi:2022pqh}
 for several recent works on BH shadow and QNMs beyond GR. }. 
 The paper is organized as follows. In Sec. \ref{Dyn}, we outline $f(T)$ teleparallel gravity introducing the concept of dynamical horizon. The violation of MSS conjecture is described in Sec. \ref{MSS}.  Chaos and geodesic instabilities are discussed in Sec. \ref{Chaos}. We consider QNMs and BH shadow  in Sec. \ref{QNM}. Conclusions are reported in Sec. \ref{Conc}.

\section{Dynamical horizon in $f(T)$ gravity}
\label{Dyn}
The $f(T)$ teleparallel gravity is the straightforward extension of Teleparallel Equivalent Gravity \cite{Carmen}. Its action reads as \cite{Cai:2015emx} 
\begin{equation}
\label{fTT}
I=\frac{1}{16\pi}\int d^4 x |e|f(T)+I_{m}
\end{equation}
(in units $G=c=1$ where $G$ is the Newton constant and $c$ is the speed of light in vacuum),
where the metric and coordinates are projected in the tetrad representation as 
$ds^2 =g_{\mu \nu}dx^{\mu} dx^{\nu}= \eta_{ij} \theta^{i} \theta^{j}$, $dx^{\mu}=e_{i}^{\mu} \theta^{i}$, 
$\theta^{i}= e_{\mu}^{i} dx^{\mu}$, $e^{\mu}_{i}e^{i}_{\nu}=\delta_{\nu}^{\mu}$,
$\sqrt{-g}=e={\rm det}[e_{\mu}^{i}]$ ($i,j=1,2,3,4$, $\mu=0,1,2,3$).
Above, we introduced $I_{m}$ as the generic action of matter sector. $T$ is the trace of torsion tensor.
The field equations  corresponding to  Action \eqref{fTT}
are 
\begin{equation}
\label{EoM}
4\pi T_{\mu}^{\nu(m)}=S_{\mu}^{\nu\rho} \partial_{\rho} T f''+ [e^{-1} e_{\mu}^{i} \partial_{\rho}(e S_{\alpha}^{\nu\rho} e_{i}^{\alpha})+T_{\mu\sigma}^{\alpha}S_{\alpha}^{\nu\sigma}]f'+\frac{1}{2}f \delta_{\mu}^{\nu}
\end{equation}
where ${\displaystyle f'=\frac{df}{dT}}$, ${\displaystyle f''=\frac{d^{2}f}{dT^{2}}}$.
Here 
\begin{equation}
\label{STTR}
T=T_{\mu\nu}^{\alpha}S^{\mu\nu}_{\alpha}\, , 
\end{equation}
is the trace  of torsion tensor 
\begin{equation}
\label{Tmm}
T_{\mu\nu}^{\alpha}=e_{i}^{\alpha}(\partial_{\mu} e_{\nu}^{i}-\partial_{\nu} e_{\mu}^{i})\, ,
\end{equation}
is the torsion tensor obtained by the Weitzenb\"ock connection 
\begin{equation}
\Gamma^{\rho}_{\mu\nu}= e_i^{\rho}\partial_{\nu}e^i_{\mu}\,.
\end{equation}
The quantity
\begin{equation}
\label{Smm}
S_{\alpha}^{\mu\nu}=\frac{1}{2}(\delta_{\alpha}^{\mu} T_{\beta}^{\nu \beta} - \delta_{\beta}^{\mu} T_{\alpha}^{\nu\beta} +K_{\alpha}^{\mu\nu})\, ,
\end{equation}
is the superpotential
with $K_{\alpha}^{\mu\nu}$ the contortion tensor. In the limit of $f''\rightarrow 0$, with prime indicating the derivative with respect $T$, GR equivalent teleparallel gravity is obtained.  See \cite{Cai:2015emx} for details on the above notations. Finally,  $T_{\mu}^{\nu(m)}$ is the standard energy-momentum tensor related to the matter sector. 

%{\bf Black Hole dynamical horizon}.

In $f(T)$ gravity, the Raychaudhuri equation 
is modified, with a deformation of the expansion, shear, vorticity and acceleration with respect to the same quantities of GR. It is:
\begin{equation}
\label{fT}
\hat{\theta}=\theta_{GR}-2T^{\rho}U_{\rho}\, , \,
\hat{\sigma}_{\mu\nu}=(\sigma_{GR})_{\mu\nu}+2h_{\mu}^{\rho}h_{\nu}^{\sigma}K_{(\rho \sigma)}^{\lambda}U_{\lambda}\, , 
\end{equation}
\begin{equation}
\label{nn}
\hat{\omega}_{\mu\nu}=(\omega_{(GR)})_{\mu\nu}+2h_{\mu}^{\rho}h_{\nu}^{\sigma}K_{[\rho \sigma]}^{\lambda}U_{\lambda}\,,
\end{equation}
\begin{equation}
\label{acc}
\hat{a}_{\rho}=(a_{GR})_{\rho}+U^{\mu}K_{\mu\rho}^{\sigma}U_{\sigma}\, ,
\end{equation}
\begin{equation}
\label{NablaU}
\tilde{\nabla}_{\mu}U_{\nu}=\hat{\sigma}_{\mu\nu}+\frac{1}{3}\hat{\theta}h_{\mu\nu}+\hat{\omega}_{\mu\nu}-\hat{a}_{\mu}U_{\nu}\, . 
\end{equation}
Not only variables but also the equation structure results deformed as:
$$\dot{\hat{\theta}}=-\frac{1}{3}\hat{\theta}^{2}+\hat{\omega}_{\mu\nu}\hat{\omega}^{\mu\nu}-\hat{\sigma}_{\mu\nu}\hat{\sigma}^{\mu\nu}-\mathcal{R}_{\mu\nu}U^{\mu}U^{\nu}-\tilde{\nabla}_{\rho} \tilde{a}^{\rho} $$
\begin{equation}
\label{eqq}
-2U^{\nu}T_{\mu\nu}^{\sigma}\Big(\frac{1}{3}h_{\sigma}^{\mu}\hat{\theta}+\hat{\omega}_{\sigma}^{\mu}+\hat{\sigma}_{\sigma}^{\mu}-U_{\sigma}\hat{a}^{\mu}\Big)\, . 
\end{equation}
with $\dot{\hat{\theta}}=d{\theta}/d\lambda$ and $\lambda$ affine parameter. 
We have introduced  the following modified Ricci tensor:
\begin{equation}
\label{ffft} \mathcal{R}_{\mu\nu}=R_{\mu\nu}-2\nabla_{\mu} T_{\nu}+\nabla_{\rho} K^{\rho}_{\mu\nu} + K_{\mu\lambda}^{\rho}K_{\nu\rho}^{\lambda}\, ,
\end{equation}
where $R_{\mu\nu}$ is the standard Ricci tensor in metric formalism. 
In the case of null oriented surfaces, 
the above equations can be rewritten using 
${\displaystyle U^{a}\equiv k^{a}=\frac{dx^{a}}{d\lambda}}$
with $k^{2}=0$,  ${\displaystyle \hat{\theta}=k^{a}_{;a}=2\frac{1}{\Sigma}\frac{d\Sigma}{d\lambda}}$. Here $\Sigma$ is a 2-surface. 

In several $f(T)$ models, 
Eq.\eqref{eqq} leads to the evolution of  marginally outer trapped (MOT) surfaces $\Sigma$, 
satisfying the condition $\hat{\theta}_{\Sigma}(\lambda=0)=0$,
 %to transmute from time-like $\partial \theta(\Sigma)/\partial n^{a}>0$ to space-like $\partial \theta(\Sigma)/\partial n^{a}<0$.
with an upper bound on their derivative as 
\begin{equation}
\label{jjaj}
\frac{d\hat{\theta}}{d\lambda}<-\mathcal{R}_{ab}k^{a}k^{b}\, , 
\end{equation}
considering also  field Eqs.\eqref{EoM}. 
Thus a surface, initially characterized by $\hat{\theta}(0)=0$,  can dynamically evolve towards
negative values of $\hat{\theta}(\lambda)$. The condition  $\mathcal{R}_{ab}k^{a}k^{b}>0$ corresponds to $\hat{\theta}(\lambda)<\hat{\theta}(\lambda=0)-K\lambda- O(\lambda^{2})$, with $K>0$. 
In the case of a BH, this means $R_{\Sigma}(\lambda)<R_{BH}(\lambda)$ for $\lambda>0$ starting from 
an initial condition $R_{\Sigma}(0)=R_{BH}(0)$
where $R_{\Sigma,BH}$ are the $\Sigma$ and BH horizon radii respectively. This means that  the surface is trapped inside the BH and it becomes space-like oriented.  The MOT surface, that at $\lambda=0$ is at the BH horizon,  becomes smaller than the BH horizon surface for $\lambda>0$. 
Such a phenomenon is related to the dynamical increasing of  BH radius. It is 
 generated by the presence of a {\it geometric fluid} coming 
from the extra gravitational degrees of freedom related to extended gravity \cite{Addazi:2017vti}. 
There are explicit analytic solutions where the BH horizon increases in time. 
This effect is completely absent in the case  $f''(T)=0$, that is when teleparallel equivalent GR is restored.
In the case of spherically symmetric critical BHs, 
there are exact solutions with a BH radius $R_{BH}(t)=R_{BH}(0)\exp(\sin h \lambda_{G}t)$ in a 
large region of parametric space \cite{Addazi:2017lat} in both diagonal and non-diagonal tetrad basis. Here
 $\lambda_{G}$ is the gravitational Lyapunov exponent. 

\begin{figure}[ht]
\centerline{ \includegraphics [width=1\columnwidth]{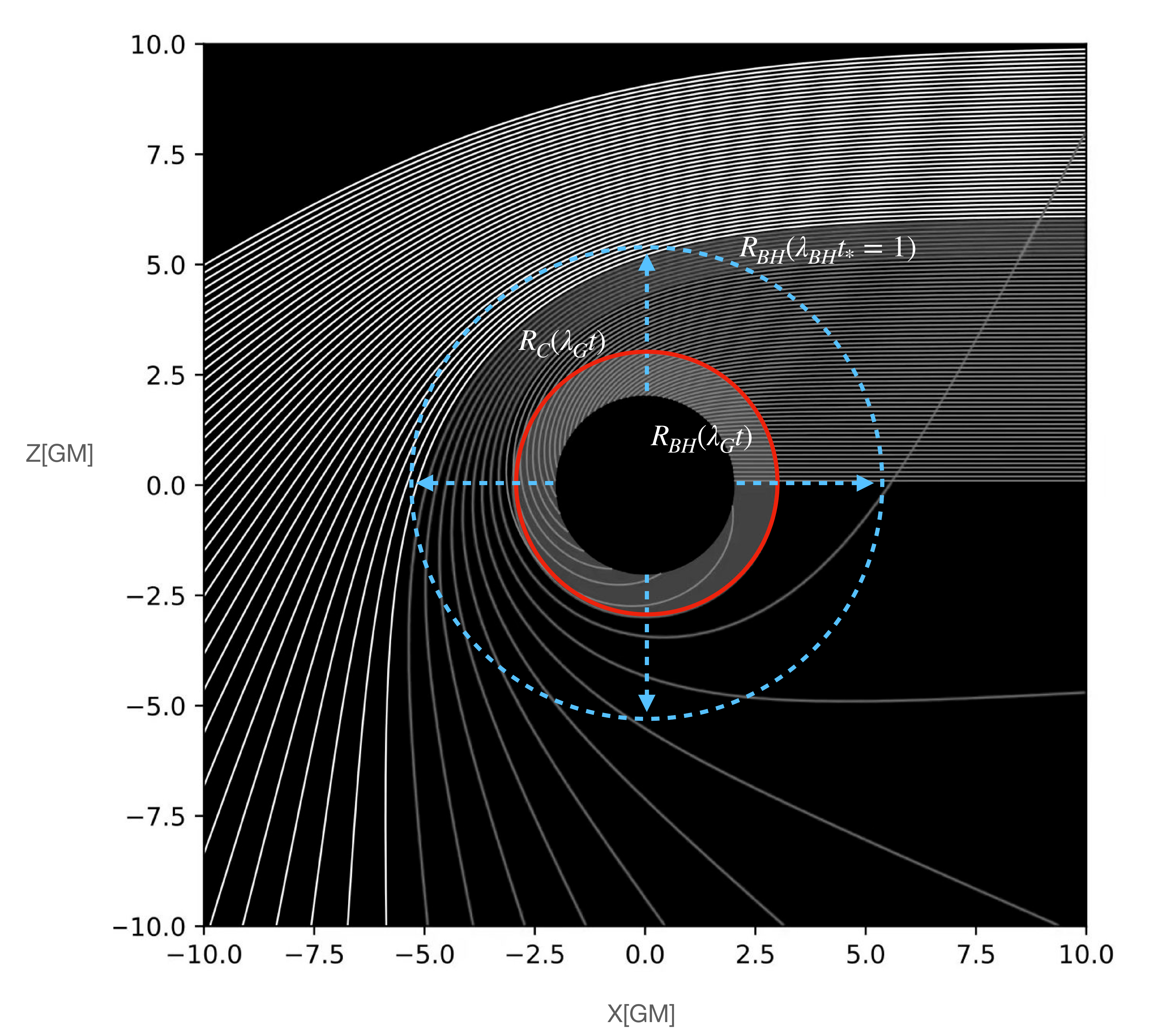}}
%\vspace*{-1ex}
\caption{Pattern of null geodesics on the  XZ plane. 
The chaotization of trajectories  around the circular radius (displayed in red) is controlled by the Lyapunov parameters $\lambda_{0},\lambda_{G}$, the ordinary and gravitational ones respectively. Indeed, the BH radius and photo-sphere critical radius $R_{BH,C}(\lambda_{G}t)$ dynamically grow in time with a rate related to $\lambda_{G}$, impacting on the stability of trajectories.  In light blue. we show the BH radius after a time transient of $t_{*}=\lambda_{G}^{-1}$ translating the geodetics (in turn not displayed at the same time $t_{*}$ for simplicity).}
%\label{fig:pep}
\end{figure}

\section{ The violation of  MSS conjecture}
\label{MSS}

The MSS conjecture states that, considering a N-body chaotic system,  commutators of two 
Hermitian operators cannot evolve faster than exponentially in time with 
 a Lyapunov parameter bounded as $\lambda \leq 2\pi \beta^{-1}$ 
(in $k_{B}=\hbar=1$ natural unit) \cite{Maldacena:2015waa}. In the case of a BH, $\beta^{-1}$ coincides with the Hawking temperature.
A typical correlator used for a diagnosis of butterfly effect in the system 
is defined as 
\begin{equation}
\label{ee}
F(t)={\rm Tr}[\zeta V(0)\zeta W(t)\zeta V(0) \zeta W(t)]\, , \,\,\, \zeta^{4}=\frac{1}{Z}e^{-\beta H}\, ,
\end{equation}
where $V,W\equiv V(t),W(t)$ are two generic OTO correlators, $H$ and $Z$ are the Hamiltonian and the partition function of the physical system respectively.
According to MSS conjecture, such an operator 
departs from a constant value $F_{d}$
after a critical scrambling time $t_{*}\sim \lambda^{-1}\log \hbar^{-1}$, 
with a growth rate that is always bounded 
as
\begin{equation}
\label{DeltF}
\frac{d}{dt}\Delta F(t)\leq \frac{2\pi}{\beta}\Delta F(t)\, , 
\end{equation}
where $\Delta F(t)=|F(t)-F_{d}|$.

The MSS conjecture is based on a simple argument 
which fails in presence of gravitational metric instabilities \cite{Addazi:2021pty}.
In this last case, 
there is a second Lyapunov exponent 
which competes with the circular trajectory instability one.
Thus, in our case, the MSS bound is modified as 
\begin{equation}
\label{aaff}
\frac{d}{dt}\Delta F\leq \lambda_{0}e^{\lambda_{G}t}(\Delta F)\, .
\end{equation}
where $\lambda_{0}\equiv \lambda(0)$.
If $\lambda_{G}>>\lambda_{0}$, 
then the  exponential can become 
larger than $1$ after a characteristic 
time $t_{g}\sim \lambda_{G}^{-1}<<\lambda(0)^{-1}$,
violating the MSS bound. 

Such a generic argument can be specifically implemented in case of classical chaotic systems, 
considering the following correlators of particle trajectories:
\begin{equation}
\label{CC}
C(t,t')=\Big\{\varphi(t),\varphi(t') \Big\}_{P.B.}=ce^{\lambda_{0} (t-t')e^{\lambda_{G}(t-t')}}\,, 
\end{equation}
where $\varphi(t)$ is the trajectory azhimutal angle variable and $\{...\}_{P.B.}$ the Poisson Brackets.

\section{ Chaos and Geodetic instabilities}
\label{Chaos}

Let us consider now a classical chaotic system, with Lyapunov parameter
related to the rate of convergence or divergence of trajectories 
inside the phase space $\{Q_{i},P_{i}\}$ with $i=1,2,3$ the configuration space-coordinate indices.
Let us consider a certain solution $\bar{Q}(t)$ of the equation of motion $\dot{Q}_{i}=P_{i}$
and let us consider a perturbation around it, that is  $\delta Q_{i}$.
The perturbation satisfies a linearized equation of motion 
$d\delta Q_{i}/d t=K_{ij}(t)\delta Q_{j}$ where $K_{ij}=\partial P_{j}/\partial Q_{i}|_{\bar{Q}(t)}$.
The integral form of the perturbation solution is $\delta Q_{i}=U_{ij}(t) \delta Q_{j}(0)$
with $\dot{U}_{ij}=K_{im}U_{mj}$. 
The principal Lyapunov parameter is operatively defined as 
\begin{equation}
\label{lambda}
\lambda=\mbox{ lim}_{t\rightarrow \infty}\frac{1}{t}\log \frac{U_{ii}(t)}{U_{ii}(0)}\, . 
\end{equation}
%which in case of a particle on a radially trajectory can be approximated as 
%$\lambda \sim \lambda_{0}e^{\lambda_{g}t}$.
In the case of circular geodesics with slow radial time-variation, 
the Lyapunov parameter has the form 
\begin{equation}
\label{lambaaa}
\lambda(t)=\frac{1}{2}\sqrt{(2g-rg')V_{r}''}\sim \lambda_{0}e^{\lambda_{G}t}\, ,  ,
\end{equation}
with the corresponding metric and 
point-like particle Lagrangian 
defined as 
\begin{equation}
\label{dss}
ds^{2}=g(r,r_{BH}(t))dt^{2}-\frac{1}{g(r,r_{BH}(t))}dr^{2}-r^{2}d\Omega^{2}\, ,
\end{equation}
\begin{equation}
\label{2L}
2\mathcal{L}=g(r,r_{BH}(t))\dot{t}^{2}-\frac{1}{g(r,r_{BH}(t))}\dot{r}^{2}-r^{2}\dot{\phi}^{2}\, .
\end{equation}
Above we defined the effective potential, derived from the Euler-Lagrange equations of Eq.\eqref{2L},
with the form 
\begin{equation}
\label{Vrdd}
V_{r}''=-2\Big(\frac{3gg'/r-2g'^{2}+gg''}{2g-rg'} \Big)\, . 
\end{equation}
Here, we performed the slow varying radii approximation
$\dot{r}_{BH},\ddot{r}_{BH}<<r_{BH}$. 
The overall picture of the geodetic dynamics is displayed in Fig.1. 

\section{Quasi-Normal modes and Black Hole Shadow}
\label{QNM}

Let us consider now a scalar field around a BH, described by 
the Klein-Gordon equation  ${\displaystyle \frac{1}{\sqrt{-g}}\partial_{\mu}(\sqrt{-g}g^{\mu\nu}\partial_{\nu} \Phi)=0}$.
In spherically symmetric background, we can decompose $\Phi$ in 
a radial and angular part as $\Phi(t,r,\theta,\phi)=\frac{1}{r}R(r,t)Y_{lm}(\theta,\phi)$ 
where $Y_{lm}$ are  spherical harmonics and 
the radial part $R$ has the following  equation of motion:
\begin{equation}
\label{ssaaa}
[\partial_{t}^{2}-\partial_{r_{*}}^{2}+V]R(r,t)=0
\end{equation}
where ${\displaystyle r_{*}=\int_{-\infty}^{\infty}\frac{dr}{f(r)}}$.
 Here $V$ is  the effective potential. 
In the limit of static background, such an equation 
reduces to the well known Regge-Wheeler one, 
with $V=g(r)[l(l+1)r^{-2}+r^{-1}g'(r)]$.
Nevertheless, in our case, the BH is 
expanding and changing in time.
Thus $V\equiv V(r,t)$,  $r_{*}\equiv r_{*}(t)$,
and $\partial_{r_{*}}=(\frac{\partial r_{*}}{\partial t})^{-1}\frac{\partial}{\partial t}$. 
On the other hand, we can consider a slow varying approximation 
where $\dot{r}_{BH},\ddot{r}_{BH}<<r_{BH}$, which is the case of antievaporation in $f(T)$ gravity. 
In such a case, 
Eq.\ref{ssaaa} has a similar structure of the Regge-Wheeler 
equation, with a modified potential 
\begin{equation}
\label{equation}
V(t,r)\simeq g(r,t)\Big[\frac{l(l+1)}{r^{2}}+\frac{1}{r}\frac{dg(r,t)}{dr} \Big]\, .
\end{equation}
If the variation of $f(r)$ in time is negligible, one can derive 
a modified QNM equation with slowly 
varying characteristic frequency $\omega(t)$:
\begin{equation}
\label{dpsi}
\frac{d^{2}\Psi(r,t)}{dr_{*}^{2}}+[\omega(t)^{2}-V(r,t)]\Psi(r,t)=0
\end{equation}
where $R(r,t)\simeq \psi(r,t)e^{-i\omega(t)t}$. 
Under this assumption, we can search for QNMs 
as saddle solutions with QNM frequencies 
which evolve in time and satisfy the condition
\begin{equation}
\label{kka}
\frac{i\omega^{2}(t)-V(r,t)}{\sqrt{-2\nabla^{2}V(r,t)}}+const=n+\frac{1}{2}\, . 
\end{equation}

%$dr_{*}/dt=f^{-1}dr/dt+d(f^{-1})dr\simeq f^{-1}dr/dt$
Following this approach, we can arrive to modified QNMs 
(MQNMs) with QNM frequencies 
\begin{equation}
\label{omegaQNM}
\omega_{QNM}(t)=\Omega_{c}(r_{BH}(t))l-i(n+1/2)|\lambda(t)|\, ,
\end{equation}
with $\Omega_{c},\lambda$ depending on the dynamical metric.
In the case of nearly Schwarzschild-de Sitter BHs or primordial BHs,
we obtain  $|\lambda|=\Omega_{c}$ and 
\begin{equation}
\label{omeagQNM}
\omega_{QNM}=\Omega_{c}(r_{BH}(t))[l-i(n+1/2)]
\end{equation}
with 
\begin{equation}
\label{aaa}
\Omega_{c}(r_{0}(t))=\frac{r_{dS}(t)-r_{BH}(t)}{2r_{BH}(t)^{2}}\simeq \epsilon \frac{1}{2}e^{c_{\pm}t}
\end{equation}
where $\epsilon\equiv r_{dS}(0)-r_{BH}(0)<<r_{BH}(0)$. Here $c_{\pm}$ are positive and negative coefficients respectively. We defined $r_{dS,BH}$ as the dS and outer BH event horizon respectively. 
In the case of $c_{-}$, $\Omega_{c}$ exponentially explodes in time, altering the QNMs. 

Other modifications on QNMs in $f(T)$-gravity have been  recently studied 
in Ref.\cite{Zhao:2022gxl} for spherically static symmetric solutions. 
In our case, we consider the leading order modifications from a dynamical 
horizon. It is worth  noticing that the two kinds of modifications introduced by $f(T)$ gravity with respect to GR can both be present in the case of spherically symmetric solutions with time-varying horizon.

\section{Discussion and Conclusions }
\label{Conc}

We showed that  the MSS bound 
can be violated in a class of BH solutions derived from $f(T)$ teleparallel  gravity. 
Such an effect was previously detected in
$f(R)$ metric gravity for several functional profiles and parameters \cite{Addazi:2021pty}. Recently, a different kind of MSS bound violation has been reported  for  charged Kiselev BHs \cite{Gao:2022ybw}. 

Specifically, we found that there is a new Lyapunov exponent controlling chaotization 
sourced by gravitational  terms. 
The gravitational Lyapunov parameter 
can be higher than the standard Lyapunov one derived from the MSS conjecture.
In particular, we have shown that in $f(T)$ teleparallel  gravity, the 
chaotization of BH systems can 
be more efficient than in GR, 
with important consequences for BH shadow physics. 
In fact, the gravitational Lyapunov constant 
not only alters the chaotic geodesics 
around the BH photo-sphere, 
but it  modifies  also the QNMs. 
In principle, such a  a violation of MSS bound in $f(T)$ teleparallel gravity 
could be tested in forthcoming  BH shadow and GW detections.
Phenomenology  and possible matching with observational data, see for example \cite{Jusufi:2022loj},  will be the argument of a forthcoming study.

\vspace{0.2 cm}

{\bf Acknowledgements}. 
A.A. work is supported by the Talent Scientific Research Program of College of Physics, Sichuan University, Grant No.1082204112427 \& the Fostering Program in Disciplines Possessing Novel Features for Natural Science of Sichuan University, Grant No.2020SCUNL209 \& 1000 Talent program of Sichuan province 2021. S.C. acknowledges the support of Istituto Nazionale di Fisica Nucleare (INFN) (iniziative specifiche MOONLIGHT2 and QGSKY). This paper is based upon work from COST Action CA21136 Addressing observational tensions in cosmology with systematics and fundamental physics (CosmoVerse) supported by COST (European Cooperation in Science and Technology).

\end{document}